\documentclass[twocolumn,showpacs,preprintnumbers,amsmath,amssymb]{revtex4}

\usepackage{graphicx}
\usepackage{dcolumn}
\usepackage{bm}
\usepackage{textcomp}
\usepackage{wasysym}
\usepackage{mathrsfs}

\begin{document}
\title{Coexistence and interconvertibility of ferromagnetic and antiferromagnetic phases in the single crystal of Mn$_3$ single-molecule magnet }
\author{Yan Cui,  Yan-Rong Li, Rui-Yuan Liu, Zhen-Yu Mi and Yun-Ping Wang}%
\affiliation{ Beijing National Laboratory for Condensed Matter
Physics, Institute of Physics, Chinese Academy of Sciences, Beijing
100190, People's Republic of China}


\date{\today}
\begin{abstract}
We report the coexistence of ferromagnetic  and antiferromagnetic
phases in the single crystal of Mn$_3$ single-molecule magnet. The
coexistent state appears within a certain period of time in the
progress of either oxidation or reduction during a reversible
oxidation-reduction process, when the sample is exposed in the air
(oxygen) or the methyl vapor. We noticed an apparent change in the
molecular structure, which is also reversible in terms of that the
methyl group is dropped or added to the molecules during the
oxidation or reduction. The absence or presence of the methyl group
in the molecules exert an essential impact upon the intermolecular
exchange interaction, and the ferromagnetic phase comes from the
heterogenous intermolecular bonds between pairs of molecules of
which one molecule has a methyl group whereas the other has lost the
methyl group. The reversible change in molecular structure suggests
the magnetic structure of Mn$_3$ might be designed and modulated at
molecular scale, which implies Mn$_3$ has a great application
potential.
\end{abstract}

\maketitle

\begin{figure}[ht]
\scalebox{0.75}{\includegraphics[bb=50 10 9cm 15cm]{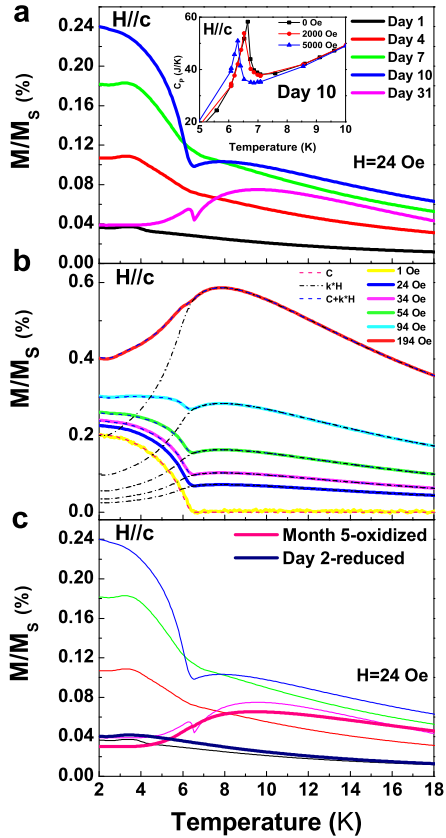}}
 \caption
{(Color online) (a), The normalized magnetization vs temperature
($M/M_s$-$T$) curves of sample \textbf{$\#1$} preserved in the air
on different days. The applied field is 24Oe. The inset shows the
heat capacity ($C_P$)
 curves of sample \textbf{$\#1$} on the 10th day at different fields. (b), The $M/M_s$-$T$ curves of sample \textbf{$\#1$}
 on the 10th day at different magnetic fields.They may be well fitted by $M(T,H) = k(T)H +C(T)$ marked by dash lines.
 (c), The $M/M_s$-$T$ curves of the fully oxidized sample \textbf{$\#3$} before and after 24-hour reduction are presented in thick lines.
 The sample \textbf{$\#3$} is fully oxidized by being preserved in the air for 5 months. The applied field is 24Oe.}
\end{figure}
 The coexistence of multiple magnetic phases has attracted considerable interest in recent years,
 due to the potential for technological applications\cite{1,2,3} and theoretical researches\cite{4,5}.
 In the reported researches, so far,
 the coexistent systems of multiple magnetic phases are
 polycrystals or crystals with apparent defects\cite{1,3,6,7,8}
 in which the system structure as well as the mechanism of the formation of
 the coexistent state are complicated, making it difficult to anticipate feasible applications.
 In this paper, for the first time, we report the coexistence and interconvertibility of ferromagnetic (FM)
 and antiferromagnetic (AFM) phases in the single crystal of Mn$_3$ single-molecule magnet (SMM)\cite{9,10,11}
 which is without apparent defect and displays quantum tunneling of magnetization (QTM) at low temperature\cite{12,13}.
 We have observed the coexistent state of FM and AFM phases during an oxidation process
 when the sample is preserved in the air or oxygen, as well as the reduction process when the oxidized sample
 is preserved in methanol gas. It is predicted that the oxidation-reduction process may be controlled by
 electrical stimuli\cite{14} of scanning-tunneling-microscopy (STM) tip, which implies the magnetic structure may be
 designed at molecular scale, suggesting Mn$_3$ may have a great potential in extensive applications of nanodevices
 for magnetic storage and spintronics\cite{15,16,17,18,19}.

 The single crystals of single-molecule magnet Mn$_3$([Mn${_3}$O(Et-sao)${_3}$(ClO${_4}$)(OH)${_3}$]) used in our experiment are
 synthesized according to the crystal growth procedures reported by Inglis et al\cite{9}.
 Mn$_3$ is known to be a SMM with AFM intermolecular exchange interaction which only exists in
 ab plane with a honey-comb structure\cite{11}, and hence may be regarded as a two-dimensional magnetic system.
The Hamiltonian of a Mn$_3$ molecule can be described as:
\begin{equation}
\hat{\mathscr{H}}=-D\hat{S}_z{^2}+g\mu_{0}\mu_{B}\textbf{H}\cdot\textbf{S}-J\sum_{i=1}^{3}\hat{S}_{iz}\hat{S}_{z},
\end{equation}
The first term represents the zero-field splitting energy, which
produces a uniaxial anisotropy barrier separating degenerate
opposite projections of the spin along the magnetic easy
axis\cite{12}. The second term is the Zeeman energy resulting from
the interaction of the spin with an applied magnetic field. The
third term is the intermolecular exchange interaction energy between
one molecule and its three neighboring molecules.

The measurements are performed using 7T SQUID-VSM (Quantum Design),
with three differently sized fresh samples preserved in the air or
oxygen. The sizes of sample \textbf{$\#1$} to \textbf{$\#3$} are
2.0mm$\times$2.0mm$\times$0.8mm, 1.8mm$\times$1.5mm$\times$0.5mm,
and 1.8mm$\times$1.0mm$\times$0.5mm, respectively. Each of the
sample is well oriented and fixed on a home-made Teflon cubic, which
is glued on the sample holder. The magnetic easy axis is ensured to
be parallel to the applied field. Fig.1a presents the normalized
magnetization vs temperature ($M/M_s$-$T$) curves of sample
\textbf{$\#1$} preserved in the air on different days. It is seen
that, FM phase starts appearing as time goes by, and becomes
apparent on Day 10. In Fig.1b, the $M/M_s$-$T$ curve at 1Oe
indicates an apparent spontaneous magnetization at Tc=6.5K shown.
The $M/M_s$-$T$ curves of sample \textbf{$\#1$} on Day 10 shown in
Fig.1b can be well fitted by $M(T,H)=k(T)H+C(T)$ at low magnetic
fields, where the field-independent term $C(T)$ is from FM phase
while the field-proportional term $k(T)H$ includes the contributions
of AFM phase and a small amount of isolated FM occurrence. The
fitting curves are indicated with the dash lines. It should be noted
that, there is no field-independent term $C(T)$ when the field is
applied perpendicular to the easy axis of the crystal while the
magnetic moment of FM phase is parallel to the easy axis of the
crystal. With the value of $C(T)$ at $T$=2K, the proportion of FM
phase can be figured out. For all samples that we have measured, the
proportion of FM phase is not more than 0.2\%. Meanwhile, AFM phase
is also observed by heat capacity measurement which is performed on
PPMS (Quantum Design). The inset in Fig.1a shows the heat capacity
vs temperature curves of sample \textbf{$\#1$} at several given
magnetic fields measured on Day 10. The peak moves to low
temperature with increasing field, indicating the typical
characteristic of AFM phase transition. Therefore, it is evident
that FM and AFM phases coexist during the oxidation process.
\begin{figure}[ht] \scalebox{0.40}{\includegraphics[bb=450 45 6cm
22.8cm]{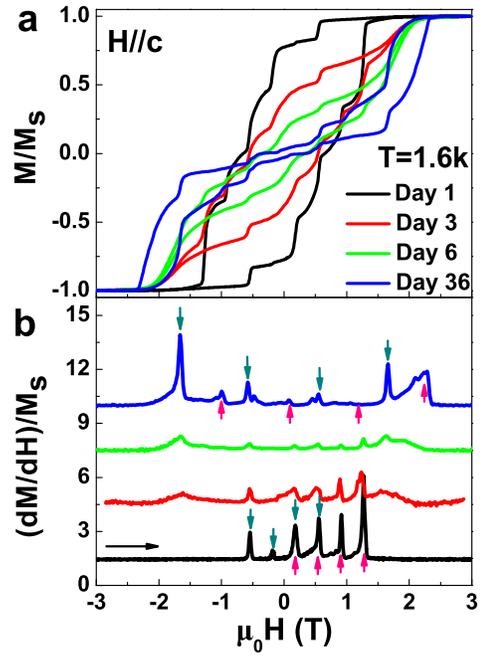}}
 \caption
{(Color online) (a), Normalized magnetization hysteresis loops
($M/M_s$-$H$) of Mn$_3$ sample \textbf{$\#2$} preserved in the air
on different days. The sweep rate is 50Oe/s. (b), Derivative curves
from -3T to 3T which are shifted along y axis for clarity. The QTM
peaks from $|$-$6$$>$ to $|6$$>$ and $|$-$6$$>$ to $|5$$>$ of
Mn$_{3}$(\uppercase\expandafter{\romannumeral1}) (black line) and
Mn$_{3}$(\uppercase\expandafter{\romannumeral2}) (blue line) are
both indicated by dark cyan arrows and pink arrows, respectively.}
\end{figure}

The normalized magnetization hysteresis loops ($M/M_s$-$H$) on
different days of sample \textbf{$\#2$} preserved in the air
measured at $T$=1.6K and the derivatives of the $M/M_s$-$H$ curves
with increasing field in the loops are presented in Fig.2. Black
loops/curves are of the fresh Mn$_3$ (hereafter
Mn$_{3}$(\uppercase\expandafter{\romannumeral1})), blue loops/curves
are of the fully oxidized Mn$_3$ (hereafter
Mn$_{3}$(\uppercase\expandafter{\romannumeral2})), red and green
loops/curves are of Mn$_3$ in the states between fresh and fully
oxidized. In Fig.2a, the hysteresis loops become narrower as time
goes by, indicating the anisotropy energy barrier is reducing. In
Fig.2b, for the derivative curves, the magnetic moment is saturated
at $H_z$=-3T, and the magnetic field is swept from -3T to 3T. It is
seen that, the first QTM peak deviates from zero magnetic field,
which is expected in the system with
intermolecular-exchange-coupling \cite{20}. The resonant field of
the QTM from $|$-$S$$>$ to $|S-l$$>$ spin state in single-molecule
magnets with identical exchange interaction can be described
as\cite{21}:
\begin{equation}
H_{z}=lD/g\mu_{0}\mu_{B}+(n_{\downarrow}-n_{\uparrow})JS/g\mu_{0}\mu_{B},
\end{equation}
where $n_{\downarrow}$ and $n_{\uparrow}$ represent the number of
a tunneling molecule's neighboring molecules which occupy the
$|$-$6$$>$ and $|6$$>$ state respectively.
\begin{figure}[ht]
\scalebox{0.21}{\includegraphics[bb=320 80 30cm 48cm]{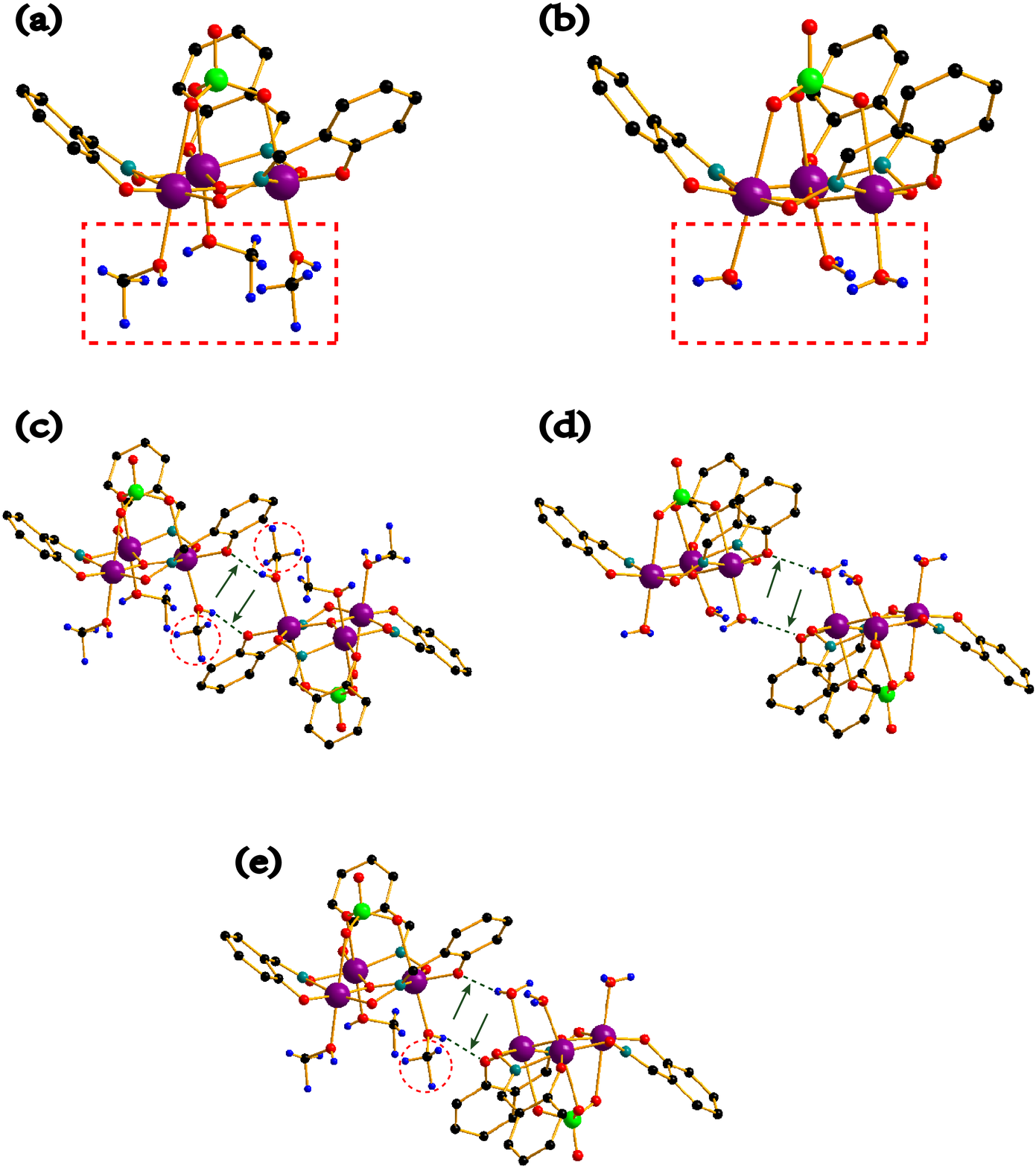}}
 \caption
{(Color online) (a), molecule structure of
Mn$_{3}$(\uppercase\expandafter{\romannumeral1}). (b), molecule
structure of Mn$_{3}$(\uppercase\expandafter{\romannumeral2}). (c),
hydrogen bonds between
Mn$_{3}$(\uppercase\expandafter{\romannumeral1}) molecules. (d),
hydrogen bonds between
Mn$_{3}$(\uppercase\expandafter{\romannumeral2}) molecules. e,
hydrogen bonds between
Mn$_{3}$(\uppercase\expandafter{\romannumeral1}) molecule and
Mn$_{3}$(\uppercase\expandafter{\romannumeral2}) molecule. Color
code: Mn, purple; N, cyan; O, red; Cl, green; C, black; H, blue. The
ethyl groups on Et-sao ligands are omitted for clarity. The hydrogen
bonds are indicated by dark green arrows. The methyl groups near the
hydrogen bonds are circled by red dashes. }
\end{figure}

Therefore, the series of QTM peaks of
Mn$_{3}$(\uppercase\expandafter{\romannumeral1}) (black line)
indicated by dark cyan arrows and pink arrows, originate from the
QTM from $|$-$6$$>$ to $|6$$>$  and $|$-$6$$>$ to $|5$$>$
respectively. These peaks are succeeded by new series of QTM peaks
(in the red, green and blue lines) as time goes by. The QTM from
$|$-$6$$>$ to $|6$$>$ and $|$-$6$$>$ to $|5$$>$ of
Mn$_{3}$(\uppercase\expandafter{\romannumeral2}) (blue line) are
indicated by dark cyan arrows and pink arrows as well. Apparently,
Mn$_{3}$(\uppercase\expandafter{\romannumeral1}) and
Mn$_{3}$(\uppercase\expandafter{\romannumeral2}) molecules are
coexistent during the oxidation process, as the QTM peaks of both
Mn$_{3}$(\uppercase\expandafter{\romannumeral1}) and
Mn$_{3}$(\uppercase\expandafter{\romannumeral2}) are observed in
both red and green curves. On the other hand, the QTM peaks in all
the curves are well defined, indicating that the easy magnetization
axes of all the molecules are all parallel to the applied field,
which means the sample is still a single crystal as the orientation
remains the same during the oxidation process. According to equation
(2), the value of anisotropy parameter $D$ and intermolecular
interaction $J$ for Mn$_{3}$(\uppercase\expandafter{\romannumeral1})
and Mn$_{3}$(\uppercase\expandafter{\romannumeral2}) may be
calculated out respectively. For
Mn$_{3}$(\uppercase\expandafter{\romannumeral1}), $D$=0.98K,
$J$=-0.041K; and for
Mn$_{3}$(\uppercase\expandafter{\romannumeral2}), $D$=0.925K,
$J$=-0.132K. The negative value of $J$ indicates the AFM exchange
interaction in both Mn$_{3}$(\uppercase\expandafter{\romannumeral1})
and Mn$_{3}$(\uppercase\expandafter{\romannumeral2}). The AFM
coupling parameter $J$ of
Mn$_{3}$(\uppercase\expandafter{\romannumeral2}) is over three times
of Mn$_{3}$(\uppercase\expandafter{\romannumeral1}), and the
coupling has contributed to the AFM phase transition observed in the
inset of Fig.1a. It should be noted that, FM phase also appears in
the same period of time during the process, but since it has a low
concentration in the sample, the QTM peaks of FM phase are not
observed.

\begin{figure}[ht] \scalebox{0.5}{\includegraphics[bb=100 -10 10cm
10cm]{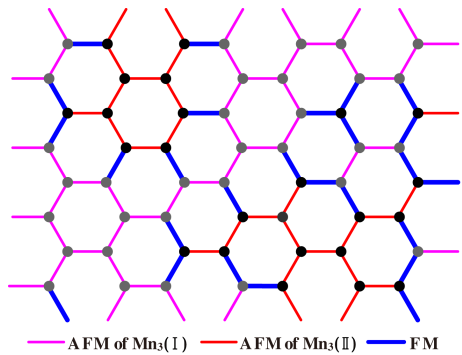}}
 \caption
{(Color online) Sketch map of magnetic structure in ab plane of
Mn$_3$ during the oxidation process. Gray dots represent
Mn$_{3}$(\uppercase\expandafter{\romannumeral1}) molecules. Black
dots represent Mn$_{3}$(\uppercase\expandafter{\romannumeral2})
molecules. The intermolecular exchange interaction only exists in ab
planes. Magenta lines represent the intermolecular bonds with AFM
exchange interaction between
Mn$_{3}$(\uppercase\expandafter{\romannumeral1}) molecules. Red
lines represent the intermolecular bonds with AFM exchange
interaction between Mn$_{3}$(\uppercase\expandafter{\romannumeral2})
molecules. Blue lines represent the intermolecular bonds with FM
exchange interaction between
Mn$_{3}$(\uppercase\expandafter{\romannumeral1}) molecules and
Mn$_{3}$(\uppercase\expandafter{\romannumeral2}) molecules.}
\end{figure}

We've also observed that, another fresh sample (sample
\textbf{$\#3$}) becomes fully oxidized after it is preserved in the
air for 5 months, with the proportion of FM phase not more than
0.001\%. Remarkably, this oxidized sample turns back to the fresh
sample when it is preserved in methanol gas (0.10bar) for more than
24 hours, which indicates the oxidation process can be reversed.
$M/M_s$-$T$ curves of sample \textbf{$\#3$} before and after the
reduction process are shown in Fig.1c. The $M/M_s$-$T$ curves of
sample \textbf{$\#3$} right after the 24-hour reduction is close to
that of sample \textbf{$\#1$} on Day 1. Since oxygen seems to play a
key role in the process, a new fresh sample is preserved in oxygen
for further test, and it is seen that, the $M/M_s$-$T$ curve on Day
6 is similar to that of sample \textbf{$\#1$} on Day 31 in Fig.1a
which is preserved in the air. The result indicates that the
oxidation process is accelerated by oxygen. Additional experiments
with the fresh samples preserved in argon or nitrogen for two weeks
and covered by AB glue for one year indicate that, the $M/M_s$-$T$
curves of these samples remains unchanged when the samples are
isolated from oxygen and methanol gas. Therefore, it is safe to
conclude that, the oxidation-reduction process may be manually
controlled by favorably applying either oxygen or methanol gas to
the sample to obtain the desired magnetic state of the sample.

In order to understand the nature of the magnetic characteristics
during this process, we have monitored the crystal structure\cite{9}
of the sample before and after the oxidation. The crystal structures
of fresh and fully oxidized samples are characterized by four-circle
X-ray diffractometer (Bruker SMART APEX-CCD) and the molecule
structures are figured out by SHELXTL. The lattice structure and
space group of the sample before and after the oxidation remain the
same. The only apparent change seen in Fig.3a and Fig.3b is that,
each Mn$_{3}$(\uppercase\expandafter{\romannumeral2}) molecule has
lost three methyl groups. The lattice constant only differs slightly
with this change. For both
Mn$_{3}$(\uppercase\expandafter{\romannumeral1}) and
Mn$_{3}$(\uppercase\expandafter{\romannumeral2}), two intermolecular
hydrogen bonds\cite{9} are formed between the oxygen and hydrogen
atoms as shown in Fig.3c and Fig.3d, and these intermolecular
hydrogen bonds determines the intermolecular exchange interaction.
The presence or absence of methyl groups results in obvious changes
of the distance and the angle of these hydrogen bonds, and hence
significantly affects the magnitude and even the sign of the
intermolecular exchange interaction parameter $J$\cite{10,20,22}.

With regard to FM phase observed in the process, it may be well
understood as the following. As shown in Fig.4, the fresh sample is
soaked in the oxygen, and hence, the
Mn$_{3}$(\uppercase\expandafter{\romannumeral1}) molecules have the
probability to be oxidized which results in the random distribution
of Mn$_{3}$(\uppercase\expandafter{\romannumeral2}) molecules. Black
dots connected by red lines represent
Mn$_{3}$(\uppercase\expandafter{\romannumeral2}) molecules with
strong AFM exchange interaction ($J$=-0.132K) which is the origin of
the AFM phase, whereas the grey dots connected by magenta lines
represent Mn$_{3}$(\uppercase\expandafter{\romannumeral1}) molecules
which are not oxidized and hence with weak AFM interaction
($J$=-0.041K). The blue lines indicate the heterogenous
intermolecular bonds (HIBs) between
Mn$_{3}$(\uppercase\expandafter{\romannumeral1}) and
Mn$_{3}$(\uppercase\expandafter{\romannumeral2}) molecules (each HIB
between the two different molecules consists of two hydrogen bonds
connecting the two molecules as shown in Fig.3e). As mentioned
above, the sample always maintains the monocrystalline structure.
Therefore, it may be convinced that the FM phase observed during the
process should come from the HIBs between
Mn$_{3}$(\uppercase\expandafter{\romannumeral1}) and
Mn$_{3}$(\uppercase\expandafter{\romannumeral2}). The presence of FM
phase indicates that the exchange interaction\cite{10,20,22} between
Mn$_{3}$(\uppercase\expandafter{\romannumeral1}) and
Mn$_{3}$(\uppercase\expandafter{\romannumeral2}) (shown in Fig.3e)
is FM interaction ($J$$>$0). In actual scenario, there are
relatively fewer HIBs with FM interaction at the beginning of the
process, these HIBs appear in small proportion, isolated and
noncoherent. With the oxidation process developing, there are more
and more HIBs appearing, and some of them will correlate with each
other by dipolar interaction to form "pieces" of FM phase which will
gradually accumulate and then the FM phase will become manifested.
Meanwhile, there are some area that most
Mn$_{3}$(\uppercase\expandafter{\romannumeral1}) molecules in it are
oxidized, hence the AFM phase is formed in these area. After that,
the FM phase will gradually diminish and disappear when more and all
molecules become oxidized eventually, and the sample will exhibit
only antiferromagnetism in the end. On the other hand, when the
oxidized sample is preserved in methanol gas, the methyl groups of
methanol will graft on
Mn$_{3}$(\uppercase\expandafter{\romannumeral2}) molecules, then a
reduction process will take place, the oxidized sample will change
back to the fresh sample.

It is very clear, the oxidation-reduction process is essentially of
that, the Mn$_{3}$ molecule loses or gains methyl groups. We have
demonstrated that the process is reversible, and may be controlled
by chemical stimuli of applying favorably oxygen or methanol gas to
the sample so that the methyl groups may be added to or dropped from
the Mn$_{3}$ molecule. It might be also possible that, the process
could be controlled by photo-irradiation\cite{23} or even electrical
stimuli \cite{14}. We are especially attracted to the thought of
electrical stimuli, with the idea of grafting the monomolecular
layer or thin film of Mn$_{3}$ on conducting substrate (the method
which has been tried for Mn$_{3}$\cite{24}), placing the device in a
mixture of oxygen and methanol gas with proper dosage, and setting a
local voltage between a STM tip and the conducting substrate to
determine oxidation or reduction process so that the intermolecular
exchange interaction may be switched between FM and AFM by tuning
the voltage, which implies the magnetic structure of Mn$_{3}$ may be
designed or modified at molecular scale. This also suggests a
possibility of extending single-molecule spintronics
device\cite{15,19} to two-dimensional spintronics device as the
magnetic coupling network may be modulated as desired.

This work was supported by the National Key Basic Research Program
of China (grant 2011CB921702).

\end{document}